\documentclass{iaus}
\usepackage{graphicx}
\newcommand\procspie{{Proc.~SPIE}}%
\newcommand\apjl{{ApJ}}%
 \newcommand\pasp{{PASP}}%
\title[Maser observations with new instruments] 
{Maser observations with new instruments}

\author[Al Wootten]   
{Alwyn Wootten$^1$
 }

\affiliation{$^1$North American ALMA Science Center, NRAO\footnote{The National Radio Astronomy Observatory is a facility of the
National Science Foundation operated under cooperative
agreement by Associated Universities, Inc.}, 520 Edgemont Rd.,
Charlottesville, Virginia 22903, USA \\ email: {\tt awootten@nrao.edu} }

\pubyear{2012}
\volume{287}  
\pagerange{1-4}
\jname{Cosmic Masers - from OH to H$_0$}
\editors{E. Humphreys \& W. Vlemmings, eds.}
\begin{document}

\maketitle

\begin{abstract}
The Atacama Large Millimeter/submillimeter Array (ALMA)\footnote{The Atacama Large Millimeter Array (ALMA) is an international astronomy 
facility. ALMA is an equal partnership between Europe and North America, 
in cooperation with the Republic of Chile, and is funded in North America 
by the U.S. National Science Foundation (NSF) in cooperation with the 
National Research Council of Canada (NRC), and in Europe by the European 
Southern Observatory (ESO) and Spain. ALMA construction and operations are 
led on behalf of North America by the National Radio Astronomy Observatory 
(NRAO), which is managed by Associated Universities, Inc. (AUI), and on 
behalf of Europe by ESO.}, and the Jansky Very Large Array (JVLA)
have recently begun probing the Universe. Both provide the largest collecting area available at locations
on a high dry site, endowing them with unparalleled potential for sensitive spectral line observations.
Over the next few years, these telescopes will be joined by other 
telescopes to provide advances in maser science, including NOEMA and the LMT.
Other instruments of note for maser science which may commence construction
include the North American Array, the CCAT, and an enlarged worldwide VLB 
network outfitted to operate into the millimeter wavelength regime.
\end{abstract}

\firstsection 
\section{Introduction}

At the last IAU Symposium on masers, construction had just begun on the Atacama Large Millimeter/Submillimeter Array (ALMA), the upgrade of the Very Large Array to the Jansky Very Large Array was being planned, and millimeter instrumentation was being introduced on the Green Bank Telescope (GBT).    CARMA had just commenced its first semester of routine science and would shortly populate its more extended arrays, enabling high resolution observations, particularly well adapted to maser observations.  The Submillimeter Array was forging new paths with high resolution observations up to the edge of the atmospheric windows near 700 GHz.  The Nobeyama Millimeter Array worked in conjunction with the 45m telescope as the Rainbow Array.  The Plateau de Bure continued to improve its bandwidth, sensitivity and resolution.  Many of the results presented at this meeting were obtained over the intervening years with these telescopes.

In this review, the status is surveyed of new instruments which will provide new science during the time to the next Symposium beyond. 

\section{Current Status}

\underline{\it ALMA}  (see \cite[Wootten and Thompson 2009]{2009IEEEP..97.1463W} for an instrument description) began its early science 'Cycle 0' operations in late September of 2011.  ALMA construction has begun its final stages; there will be at least one further Early Science call, 'Cycle 1'.  

{\underline{\it ALMA Construction}} continues for somewhat more than a year.  

At this writing, ALMA has transported 34 antennas delivered from the three vendors to the Array Operation Site (AOS), half of the final complement of 66 telescopes. At any given time the complement of antennas at the AOS is somewhat less than this as antennas return to the Operations Support Facility (OSF) for periodic maintenance, others are involved in commissioning or in the distinct Atacama Compact Array (ACA) or 12 seven meter antennas and 4 twleve meter antennas.  

Three Front End Integration Centers, located in Charlottesville, near Oxford and Taipei assembled, tested and delivered 25 additional new Front Ends, or receiver packages, during 2011 to outfit those antennas (\cite [see, for example, Ediss et al. 2010]{2010JIMTW..31.1182E}). The units in the antenna cabin through which the Front Ends communicate with the equipment in the Array Operations Site Technical Building (AOS TB) are called Antenna Articles. Sixty-six of these were delivered by the end of the year, one for each of the planned ALMA antennas. 

The Central Local Oscillator was also delivered and installed during 2011. It forms an important part of the ALMA nervous system, distributing and synchronizing signals across the array.  A major technical problem of the ALMA Local Oscillator (LO) system is to maintain accurate phase stability in the signals used in the receiver components. The central reference part of the LO system is located at the AOS in the Technical Building. From here the signal is used to lock electronic oscillators and derived frequencies in antennas located at distances of many miles across the site via optical fiber. The outgoing frequency is transmitted as the difference between two laser frequencies in the infrared. The required phase stability of the LO signal is equivalent to less than 12 femtoseconds of time, less than 38 femtoseconds of phase noise. This time equivalent phase stability is to a second as a second is to 2.6 million years!  The CLOA is now capable of supporting the 66 antennas and four subarrays built for the ALMA construction project; it may be expanded to provide signals for up to six subarrays and eighty antennas.

The antenna stations for the central cluster are in the final stages of preparation, ready for the deployment of the Early Science Array into its extended configuration early in 2012. Thus far, ALMA has operated on a network of generators. The Permanent Power System erection was completed in 2011 and is due for deployment on the array during early 2012, improving the quality and reliability of the array electrical power.
\begin{table}
\begin{center}
\caption{{\bf  Front Ends on all Antennas$^a$}}
\begin{tabular}{llllll}
\hline
     {\bf Band}&       {\bf $\lambda$}&  {\bf $\nu$ }&  {\bf Type } &T$_{rx}^b$ &  {\bf Continuum }   \\
 &{\bf mm} &{\bf GHz} & &  & {\bf Sensitivity (mJy)}$^c$ \\  
\hline
\hline
Early Science Cycles 0-1&&& & &\\
Band 3 &3& 84-116  &2SB & 45K$^b$ & 0.04 \\
Band 6 &1.3& 211-275 & 2SB &55K$^b$ & 0.11\\
Band 7 &0.9& 275-373  &2SB & 75K$^b$  & 0.19  \\
Band 9 &0.45&602-720  & DSB &110K (DSB))$^b$ & 1.0  \\
\hline
Full Operations & &&& &\\
Band 4 &2& 125-169  &2SB&$<$51K & 0.06  \\
Band 5 &1.5&163-211 & 2SB & $<$65K &  \\
Band 8 &0.6& 385-500  &2SB&$<$196K & 0.70  \\
Band 10 &0.35& 787-950  &DSB& $<$230K DSB & 0.70  \\
\hline
Water Vapor Radiometer& 1.6 &183 &DSB Schottky& ... \\
\hline
\end{tabular}
\end{center}
\noindent
$^a$ All have dual polarization with noise performance limited by atmosphere. 
$^b$  See the ALMA sensitivity calculator for T$_{sys}$ under typical atmospheric conditions.  For Early Science, actual realized receiver temperatures are given, often much better than the specification.  T$_{rx}$ for full operations is the specification value.  $^c$ For sensitivity, 50 antennas are assumed in all cases.
$^c$ 

\end{table}
\noindent

Production lines for ALMA components are shutting down as final components are delivered. The last of the four correlator quadrants was accepted. Two correlator quadrants were deployed at the AOS TB and a third was installed, while the fourth remained in Charlottesville for software testing.  That quadrant will be installed by September 2012, when correlator installation and verification will finish.  Receiver cartridges for the 3mm (ALMA Band 3) and 0.45mm (ALMA Band 9) frequencies have also completed production. See Table 1 for a current receiver summary.  Steel fabrication for the Vertex antennas was complete; the final antenna will be shipped in April 2012 for delivery a few months later. The final third of the antennas constructed by the AEM consortium are nearing completion; most of the antennas are at the ALMA site.  All sixteen antennas for the Atacama Compact Array (ACA) were delivered to Chile from Japan. The ACA with its initial complement of antennas employed its correlator to produce the first interferometric and total power data. The penultimate software release was deployed on the array and new releases of CASA software were released to the community for the further reduction of the delivered ALMA data.

During commissioning, carried out by a team of astronomers led by Project Scientist Richard Hills and his deputy Alison Peck (\cite[Hills et al. 2010]{2010SPIE.7733E..34H}, \cite [Mauersberger et al.. 2011]{2011IAUS..280P.402M}), a typical early array consisted of a few antennas tightly packed onto several of the antenna stations designated to eventually host the Atacama Compact Array (ACA) 7m antennas, as this small set of stations could be most quickly outfitted to host the few antennas in the earliest array.  This commissioning array observed maser sources often, as these bright narrowband point emission sources provide excellent targets for assessment of the delays, pointing, baselines and a host of other instrumental qualities.  The Orion SiO J=1-0 masers, observable in most weather conditions and well-placed during the first commissioning periods in early 2010, were most often observed to provide test data to evaluate the performance of the array.  As the year pressed on and weather improved, some early maser observations included 325 GHz observations of VY CMa with 3-5 antennas.  An analysis by A. Richards of this test data showed a good correspondence between these masers and those mapped at 22 GHz by other arrays, to within the limited imaging ability of the small array.  As the array grew further, to eight antennas by October, other masers were observed to provide test data, including the 658 GHz water masers toward a number of stars.  With the increase in imaging capabilities of the array, testing moved forward to the imaging of more complex objects.  NGC253 was imaged in all of the ALMA bands, as was the Beta Pictoris disk and other targets.   The focus of the commissioning team moved on to calibration and verification of ALMA results through a comparison with existing data from other arrays.  A call for science verification targets provided a favorable response, with over a hundred suggestions received.

A goal of science verification was the release of representative datasets to the community for inspection by investigators prior to the beginning of Early Science.  Several science verification datasets were released on the ALMA portal to demonstrate the quality of ALMA data and its consistency with previous observations. Eight science verification datasets and one datum have now been released.  Several publications based on Science Verification data have been published, though none has yet dealt with maser science (\cite[Herrera et al 2012]{2012A&A...538L...9H}, \cite[Oberg et al. 2012]{2012arXiv1202.3992O}).  The datasets released so far include:
\begin{itemize}
\item  TW Hya: Band 7, high spectral resolution (casaguide). Band 3, Band 6.
\item  NGC3256: Band 3, low spectral resolution (casaguide). 
\item  Antennae galaxies: Band 7, high spectral resolution, mosaic (casaguide).Ê Band 6.
\item  M100 Band 3, low spectral resolution. 
\item SgrA* Band 6, recombination lines. 
\item Proof of Concept of Response to Targets of Opportunity: The GRB 110715A. 
\end{itemize}
As commissioning continues, several additional science verification datasets will be released to demonstrate additional capabilities.

{\underline{\it ALMA Operations}} has begun with Early Science as the array grew to 16 elements (\cite [Nyman et al.(2010)]{2010SPIE.7737E...9N}, \cite[Rawlings et al.(2010)]{2010SPIE.7737E..29R}).  The Early Science Cycle 0 Call for Proposals resulted in 2898 registrations at the ALMA portal. 919 proposals were received for the 2011 June 30 deadline.  Investigators were  notified in September of the status of their proposals; the limited time available was heavily ovsersubscribed.  Scheduling and execution of the first batch of the 112 highest ranked proposals began on 30 September. By early December, quality assurance and packaging of the first project datasets had finished and delivery of the first data packages to principal investigators followed. By the February 2012 shutdown for implementation of ALMA's permanent power supply, nine periods of Early Science had been executed on the growing array.

ALMA construction will continue through 2012, which will see delivery of most of the remaining hardware. The most exciting prospect for 2012 is, of course, the expected publication of the first papers from Early Science. Some early results are making the rounds of science meetings already, whetting astronomersÕ appetites for the announcement of the Cycle 1 Call for Proposals, expected within the next few months.  During that period, of course, some of the Early Science observations will emerge into the literature,  

By 2014 ALMA will be in full operation, having begun transforming science in 2011. Having invested $\sim$\$1.3B to realize the biggest advance ever in groundbased astronomy, it is vital to plan to keep the facility upgraded to maintain and expand its capabilities.  When ALMA commenced its program of Early Science, it already eclipsed any other millimeter/submillimeter array in its sensitivity and resolution.   The ALMA Operations Plan envisaged an ongoing program of development and upgrade.   ALMA's design allows for expansion of the 50 antennas in the 12m Array to a complement of 64.  ALMA's wavelength coverage may be extended to cover 1cm to 200 $\mu$m, or a factor of 50 and an increase of more than 50\% from its first light capability.
With a modest investment of less than 1\% of capital cost per year divided among the three funding entities ALMA will lead astronomical research through the 2010 decade and beyond.   Several programs have been identified by the scientific community which spearhead a development plan.   The ALMA Development Plan will be coordinated by the Joint ALMA Observatory (JAO), which will issue calls for ALMA upgrades whose implementation will be assigned on a competitive basis. The details of this process are currently under discussion. 

Fully operational ALMA will provide beamsizes of 0".01 at 300 GHz, appropriate for the sizes of some interstellar shock features seen in for example, water maser emission (see image in \cite[Wootten et al.(2002)]{2002IAUS..206..100W}).  For a putative 1 kms$^{-1}$ spectral resolution in an hour, the brightness temperature sensitivity of ALMA is a few K, allowing sensitive spatially and spectrally resolved observations of shock structures.  The shocks shown in the image cited extend for several AU, which is several beams for nearby maser regions.
 
 {\underline{\it GBT}} (\cite[Prestage et al.(2009)]{2009IEEEP..97.1382P}, \cite[Hunter et al.(2011)]{2011PASP..123.1087H}) brings the largest monolithic and steerable surface area on the planet to provide sensitive probes of distant weak masers.  A particularly strong program there uses water masers to provide distances to maser host galaxies (see articles by Humphreys, Impellizzeri and Henkel in this volume).  The first tests of a new 68-93 GHz receiver have proved promising; a multibeam array receiver may be implemented in the millimeter bands in the near future.  Addition of the large collecting area to a sensitive VLB network will allow proper motions of masers in nearby galaxies to be measured, monitoring the kinematics of those spiral disks.  

{\underline{\it JVLA}} (\cite[Perley et al.(2009)]{2009IEEEP..97.1448P}, \cite[Perley et al.(2011)]{2011ApJ...739L...1P})  has begun providing exciting science results as demonstrated in data presented by many at this Symposium even before it reaches its full upgraded potential.  The several order of magnitude increase in correlator capacity promises delivery of sensitive spectroscopy at high spectral and spatial resolution.  As masers are often shock phenomena, investigations of the physical and chemical characteristics of the shocks will provide new insights into the origins of the masers.  In the largest configuration, the JVLA provides tenth arcsecond resolution with a brightness temperature sensitivity of a few hundred Kelvin in an hour at a spectral resolution of 1 kms$^{-1}$, appropriate to most maser emission environments.
 
\section{Next Symposium Instruments}

{\underline{\it NOEMA}}, the Northern Extended Millimeter Array at the Plateau de Bure(\cite[NOEMA Project]{}, was agreed by the IRAM Executive Council to be expanded to four additional antennas in a first phase to be completed by 2016.  A further two antennas are envisioned for construction in Phase II.  The partners agreed to extend their successful participation in IRAM through 2024.  The expansion will increase the collecting area of the array by 67\% and imaging quality by 300\% through tripling of the available baselines.   Resolution is expected to double with the extension of baselines to 1.6 km.  Continuum sensitivity will be further enhanced through a planned quadrupling of the bandwidth to 32 GHz.  The enhanced array will provide a sensitive northern hemisphere baseline for intercontintental VLBI.

{\underline{\it LMT}}, the Large Millimeter Telescope (\cite[Hughes et al 2010]{2010SPIE.7733E..31H}), began operations at its 4600m  mountaintop, Sierra Negra, in Mexico within the past year.  Initially outfitted with a 32 meter surface, expansion to the full 50m aperture is expected soon.  The antenna surface will eventually be set to an accuracy sufficient to support operations in the 300 GHz spectral window. The antenna's aperture and location make it an important element of any worldwide VLB array.

 \section{Future}

{\underline{\it The North American Array}}   (NAA) is a new initiative aimed at the
realization of the high-frequency component of the Square Kilometer
Array (SKA) program within North America.  The primary activities of the
NAA in the next decade are a technical development program and a prototype
antenna station, leading up to SKA-high construction sometime after
2020.  The key science drivers identified for the NAA cover a broad
range of modern astrophysics, from studying the formation and
evolution of planets, stars, and galaxies, to probing the overall
structure of the Universe near and far.  The primary technical goal of
NAA is to ultimately exceed the sensitivity of the JVLA by a factor of
10 or greater, and thus survey speed by 100 or more, covering at 
least the core frequency range 3--45~GHz.  
The SKA-high is envisioned as the final component
of the low/medium/high frequency triad of the International SKA
Program.

{\underline{\it CCAT}} (\cite[Radford et al.(2009)]{2009ASPC..417..113R}) will be a 25m diameter telescope for submillimeter astronomy on a
Cerro Chajnantor centrally located on the ALMA site.  Primarily focussed on submillimeter continuum survey work, it is designed to reach frequencies above the ALMA Band 10 window, which stops at 950 GHz.  With a 6" beam at that frequency, it would be useful for submillimeter maser work.  The telescope may be incorporated within the ALMA array at some future date for particular experiments, providing sensitive superterahertz capabilities.

{\underline{\it Event Horizon Telescope}  (EHT, \cite[Fish \& Doeleman(2010)]{2010IAUS..261..271F}) is aimed primarily at observations of the black holes at the center of M87 and the Milky Way, but it could also provide exceedingly high resolution images of bright masers.   Incorporating the phased ALMA into its network, sensitive measurements of SiO or other masers in nearby galaxies, importantly the Magellanic Clouds, might be made which could determine the dynamics of those galaxies.

\end{document}